\documentclass[a4paper]{article}

\usepackage[latin1]{inputenc}
\usepackage{amsfonts}
\usepackage{amsmath}
\usepackage{amssymb}
\usepackage[UKenglish]{babel}
\usepackage{fontenc}
\usepackage{graphicx}
\usepackage{amscd}
\usepackage{xypic}
\usepackage{psfrag}

\newcommand{\CC}{\ensuremath{\mathbb{C}}}
\newcommand{\RR}{\ensuremath{\mathbb{R}}}
\newcommand{\NN}{\ensuremath{\mathbb{N}}}
\newcommand{\ZZ}{\ensuremath{\mathbb{Z}}}

\renewcommand{\d}{\ensuremath{\mathrm{d}}}

\newcommand{\modsq}[1]{\ensuremath{\left|#1\right|^2}}
\newcommand{\md}[1]{\ensuremath{\left|#1\right|}}

\newcommand{\Ob}[1]{\ensuremath{\mathcal{O}(#1)}}

\newcommand{\epsi}{\varepsilon}

\newcommand{\SU}[1]{\ensuremath{\mathrm{SU}(#1)}}
\newcommand{\SL}[1]{\ensuremath{\mathrm{SL}(#1)}}

\newcommand{\su}[1]{\ensuremath{\mathfrak{su}(#1)}}
\newcommand{\sli}[1]{\ensuremath{\mathfrak{sl}(#1)}}
\newcommand{\U}[1]{\ensuremath{\mathrm{U}(#1)}}
\renewcommand{\u}[1]{\ensuremath{\mathfrak{u}(#1)}}

\newcommand{\Endo}[1]{\ensuremath{\mathrm{End}(#1)}}
\newcommand{\Hom}[1]{\ensuremath{\mathrm{Hom}(#1)}}
\newcommand{\rk}{\ensuremath{\mathrm{rank}}}

\newcommand{\im}{\ensuremath{\mathrm{i}}}
\newcommand{\half}{\ensuremath{\frac{1}{2}}}

\newcommand{\ihalf}{\ensuremath{\frac{\im}{2}}}
\newcommand{\w}{\wedge}

\newcommand{\e}{\ensuremath{\mathrm{e}}}

\renewcommand{\det}[1]{\ensuremath{\mathrm{det}(#1)}}
\newcommand{\tr}[1]{\ensuremath{\mathrm{tr} \! \left( #1 \right)} }
\newcommand{\Ad}[1]{\ensuremath{\mathrm{Ad}(#1)}}
\newcommand{\ad}[1]{\ensuremath{\mathrm{ad}(#1)}}

\newcommand{\dvol}{\ensuremath{\mathrm{dvol}}}

\newcommand{\al}{\alpha}
\newcommand{\be}{\beta}

\renewcommand{\th}{\theta}

\newcommand{\curlA}{\ensuremath{\mathcal{A}}}
\newcommand{\curlF}{\ensuremath{\mathcal{F}}}

\newcommand{\curlE}{\ensuremath{\mathcal{E}}}
\newcommand{\curlH}{\ensuremath{\mathcal{H}}}

\newcommand{\curlG}{\ensuremath{\mathcal{G}}}

\newcommand{\id}{\ensuremath{\mathrm{id}}}

\newcommand{\mfk}{\mathfrak}
\newcommand{\pd}{\ensuremath{\partial}}

\newcommand{\dgr}{\dagger}

\begin{document}

\pagestyle{plain}

\title{
\vskip -70pt
\begin{flushright}
{\normalsize DAMTP-2010-117} \\
\end{flushright}
\vskip 50pt
{\bf \Large Geometry and Energy of Non-abelian Vortices}
\vskip 30pt
}

\author{
Nicholas S. Manton\footnote{N.S.Manton@damtp.cam.ac.uk} 
\,\,\,and\,\,\,Norman A. Rink\footnote{N.A.Rink@damtp.cam.ac.uk} \\ \\
{\sl Department of Applied Mathematics and Theoretical Physics,}\\
{\sl University of Cambridge,}\\
{\sl Wilberforce Road, Cambridge CB3 0WA, England.}\\
}

\vskip 20pt
\date{December 2010}
\vskip 20pt

\maketitle

\begin{abstract}
We study pure Yang--Mills theory on $\Sigma\times S^2$, where $\Sigma$ is a compact Riemann surface, and invariance is assumed under rotations of $S^2$. It is well known that the self-duality equations in this set-up reduce to vortex equations on $\Sigma$. If the Yang--Mills gauge group is $\SU{2}$, the Bogomolny vortex equations of the abelian Higgs model are obtained. For larger gauge groups one generally finds vortex equations involving several matrix-valued Higgs fields. Here we focus on Yang--Mills theory with gauge group $\SU{N}/\ZZ_N$ and a special reduction which yields only one non-abelian Higgs field.  

One of the new features of this reduction is the fact that while the instanton number of the theory in four dimensions is generally fractional with denominator $N$, we still obtain an integral vortex number in the reduced theory. We clarify the relation between these two topological charges at a bundle geometric level. Another striking feature is the emergence of non-trivial lower and upper bounds for the energy of the reduced theory on $\Sigma$. These bounds are proportional to the area of $\Sigma$.

We give special solutions of the theory on $\Sigma$ by embedding solutions of the abelian Higgs model into the non-abelian theory, and we relate our work to the language of quiver bundles, which has recently proved fruitful in the study of dimensional reduction of Yang--Mills theory. 
\end{abstract}

\newpage

\section{Introduction}

It was first noted in \cite{Witten} that rotationally invariant instantons in Yang--Mills theory can be interpreted as vortices in lower dimensions. This reduction was originally carried out for Yang--Mills theory with gauge group $\SU{2}$ and yielded the abelian Higgs model. In recent years, more general reductions of Yang--Mills theory on spaces of the form $\Sigma\times S^2$ have been studied, where invariance under rotations of the sphere $S^2$ was assumed, see \cite{Popov_quiver, Popov_nonab, DolanSzabo, MantonSakai} and references therein. This invariant set-up generally leads to several matrix-valued Higgs fields on $\Sigma$, and the precise number and shape of the Higgs fields is determined by the Yang--Mills gauge group and the specific way in which the rotational symmetry is implemented in the theory. For example, it was shown in \cite{MantonSakai} that if the Yang--Mills gauge group is $\SU{N}$, with $N=2m$, and one chooses a symmetry reduction which breaks this group to $\mathrm{S}(\U{m}\times\U{m})$, then a single non-abelian Higgs field, which is a square $(m\!\times\!m)$-matrix, is obtained. Here we show how a non-square Higgs field arises in the reduced theory on $\Sigma$ when the Yang--Mills gauge group, assumed to be locally the same as $\SU{N}$, is broken by the rotational symmetry to a group which is locally the same as $\mathrm{S}(\U{m}\times\U{n})$, where $N=m+n$. The Higgs field is now an $(n\!\times\!m)$-matrix. We focus on the case where $\Sigma$ is a closed, compact Riemann surface, in order that the reduced theory on $\Sigma$ can have vortex solutions of finite energy.

This symmetry breaking pattern is desirable if one wants to construct the Standard Model on $\Sigma$ from a unified theory in higher dimensions. For the electroweak sector, for example, one should take $N=3$, $m=2$, $n=1$. From a purely two-dimensional point of view, the electroweak model on $\Sigma$ was studied by Bimonte and Lozano in \cite{BimonteLozano}, where $\Sigma$ was taken to be a flat torus, with euclidean signature. In this setting, a Bogomolny-type argument can be carried out on the energy of the electroweak model, and the resulting Bogomolny equations were derived in \cite{BimonteLozano}. Vortex solutions to these equations were also obtained, related to vortex solutions studied earlier in \cite{Hindmarsh, Vachaspati}. Another result of \cite{BimonteLozano} was a lower bound on the energy of the electroweak model on $\Sigma$. This bound is proportional to the area of $\Sigma$. Here we generalize the results of \cite{BimonteLozano} to arbitrary $N$, $m$, and $n$, by viewing the theory as dimensionally reduced Yang--Mills theory on $\Sigma\times S^2$. We also obtain lower and upper bounds on the energy in this generalized setting. An interpretation of the lower bound is given in terms of the vacuum structure of the Yang--Mills theory in four dimensions. It should be noted that dimensional reduction of Yang--Mills theory on $\RR^{1,3}\times S^2$ to the electroweak model on $\RR^{1,3}$ was already carried out in \cite{Manton_6d}, and we use very closely related methods here to facilitate the reduction.

The crucial point at an early stage in our analysis is the observation that the geometry of $S^2$ forces us to start with Yang--Mills theory on $\Sigma\times S^2$ with gauge group $\SU{N}/\ZZ_N$, i.e.~the quotient of $\SU{N}$ by its centre $\ZZ_N$. This impacts on the bundle structures associated with the Yang--Mills theory and the reduced theory on $\Sigma$. Most notably, it is no longer natural to think of Yang--Mills theory as being defined on a vector bundle over $\Sigma\times S^2$ since there is no rank $N$ vector bundle with structure group $\SU{N}/\ZZ_N$. Instead we introduce a principal bundle with structure group $\SU{N}/\ZZ_N$, and we regard the gauge potential of Yang--Mills theory as a connection on this principal bundle. As a consequence, the instanton number (as conventionally normalized for a gauge potential on a rank $N$ vector bundle) need no longer be an integer but is generally a fraction with denominator not bigger than $N$. The Higgs field in the reduced theory can still be regarded as a section of a vector bundle over $\Sigma$, and so the associated vortex number is integral. This vector bundle over $\Sigma$, however, is not necessarily the bundle of homomorphisms between two distinct vector bundles as in the literature \cite{GP_invariant, Alvarez1, Popov_quiver, DolanSzabo}.

In much of the recent literature on non-abelian vortices, see \cite{HanTong, Konishi, SakaiTong, EtoSakai, Baptista_nonab} and references therein, Higgs fields are taken to be matrices whose columns are charged under the gauge group, and different columns represent different flavours. Then, in addition to gauge symmetry, there is also flavour symmetry and the corresponding symmetry group acts on the Higgs field on the right. By contrast, although the non-abelian Higgs field in the theory we study is generally matrix-valued and acted on by symmetry groups from the left and the right, both group actions are gauged and neither is a flavour symmetry. Models containing several flavours are usually obtained from supersymmetric field theories by truncating these to their bosonic parts. Here we will not consider supersymmetric models; nevertheless fermions can consistently be added to invariant Yang--Mills theory on $\Sigma\times S^2$, as was done in \cite{Manton_fermions, KapeZoup, DolanSzabo}.

This paper is organized as follows. In section \ref{sec:ansatz} we review the most general ansatz for the Yang--Mills gauge potential on $\Sigma\times S^2$ that is invariant under rotations of $S^2$. Alongside of this we clarify which bundle structures are relevant in the invariant Yang--Mills theory and in the reduced theory on $\Sigma$. In section \ref{sec:reduction} we specialize to the Yang--Mills gauge group $\SU{N}/\ZZ_N$ and choose a particular class of symmetry reductions which lead to a single Higgs field on $\Sigma$ with an associated vortex number. Section \ref{sec:actions} is dedicated to reducing the Yang--Mills action and the self-duality equations in four dimensions to the energy and Bogomolny-type equations in two dimensions. We also find the relation between the topological charges in four and two dimensions, the instanton and vortex numbers. A first lower bound on the energy of the reduced theory on $\Sigma$ is obtained, and we comment on the implications of this bound for the existence of invariant vacua in the Yang--Mills theory. A sharper lower bound on the energy of solutions to the Bogomolny equations as well as an upper bound are derived in section \ref{sec:bounds}, and we present a special class of solutions to the Bogomolny equations in section \ref{sec:max_abelian}. In section \ref{sec:examples} we explicitly connect our work with \cite{BimonteLozano}, and we comment on the allowed energy ranges for $N=3$ and $N=5$. In section \ref{sec:quivers} we formulate the bundle theoretic features of our dimensional reduction scheme in the language of quivers of vector bundles. Section \ref{sec:conclusions} sums up our conclusions.

\section{Invariant Yang--Mills theory and bundles} \label{sec:ansatz}

Throughout this paper $\Sigma$ is assumed to be a closed, compact Riemann surface\footnote{Many results, especially in the present section and the next, generalize to rather arbitrary manifolds $\Sigma$.} with local complex coordinate $z$. On the sphere $S^2$ we take the complex coordinate $y$ obtained by stereographic projection. We also introduce real coordinates $x^1$, $x^2$ on $\Sigma$ and $x^3$, $x^4$ on $S^2$ by the relations
\begin{align}
 z = x^1 + \im x^2, \quad
 y = x^3 + \im x^4. \label{eq:real_coords}
\end{align}
For the metric on $M = \Sigma\times S^2$ we adopt the conventions of \cite{MantonSakai}, i.e.
\begin{align}
 \d{s^2} = \sigma(z,\bar z) \d{z}\d{\bar z} + \frac{8}{(1+y\bar y)^2} \d{y}\d{\bar y}, \label{eq:4d_metric}
\end{align}
where $\sigma$ is the conformal factor on $\Sigma$ and the second term renders $S^2$ a sphere of radius $\sqrt{2}$ with Gauss curvature $\half$. The corresponding volume forms on $\Sigma$ and $S^2$ are
\begin{align}
 \dvol_{\Sigma} = \sigma\, \d{x^1}\w\d{x^2}, \quad
 \dvol_{S^2} = \frac{8}{(1+y\bar y)^2}\, \d{x^3}\w\d{x^4},
\end{align}
and the area of $\Sigma$ is denoted by $A_{\Sigma}$.

We consider pure Yang--Mills theory on the product space $M = \Sigma\times S^2$, and regard it as a theory of a connection $\omega$ defined on a principal bundle $P$ over $M$. The gauge potential $\curlA$ is obtained from $\omega$ by means of a local section $s\colon U\to P$, $U\subset M$ open,
\begin{align}
 \curlA = s^*\omega,
\end{align}
where the right hand side denotes the pull-back of $\omega$ under $s$. For our purposes it is best to regard the sphere as the coset space $S^2 = \SU{2}/\U{1}$. This introduces a natural transitive action of $\SU{2}$ on the sphere $S^2$, and this action extends to $M$ by acting trivially on $\Sigma$. We can then consider $\SU{2}$-equivariant principal bundles over $M$ and $\SU{2}$-invariant connections on them. Phrased in a less technical fashion, we are interested in $\SU{2}$-invariant Yang--Mills theory on $\Sigma\times S^2$. From the point of view of the surface $\Sigma$ this amounts to dimensional reduction of Yang--Mills theory on $\Sigma\times S^2$, where the sphere $S^2$ is treated as an internal space.

The goal of the present section is to identify the geometric structures on $\Sigma$ that arise from the reduction of $\SU{2}$-invariant Yang--Mills theory. The tools we are going to use are the results of the analysis in \cite{Harnad}, which generalize Wang's theorem \cite{Wang}. Similar treatments which by-pass the analysis of connections and principal bundles by focusing on the gauge potential only are \cite{ForgacsManton, KapeZoup}, and their approach is usually referred to as \textit{coset space dimensional reduction}.  

First recall (from \cite{Harnad} for example) that every $\SU{2}$-equivariant principal bundle over $S^2$ with structure group $\curlG$ is isomorphic to a quotient space $P_\lambda$ defined by 
\begin{align}
 P_{\lambda} = \SU{2}\times_\lambda \curlG,
\end{align}
where elements in $\SU{2}\times \curlG$ are identified by
\begin{align}
 (S,g) \sim (S S_0, \lambda(S_0)^{-1}g), \quad S_0 \in\U{1},
\end{align}
and $\lambda$ is a homomorphism $\lambda\colon \U{1} \to \curlG$. 
The projection map $\pi\colon P_\lambda \to S^2$ is given by
\begin{align}
 (S,g) \mapsto [S], 
\end{align}
where $g\in\curlG$ and $[S]$ denotes the left-coset $\{S\cdot\U{1}\}$ in $\SU{2}$. Note that the $P_{\lambda}$ are isomorphic for different $\lambda\colon\U{1}\to\curlG$ within the same conjugacy class.

Now let $P$ be an $\SU{2}$-equivariant principal bundle on $\Sigma\times S^2$ and choose an open covering $\{U_i\}_{i\in I}$ of $\Sigma$ such that all $U_i$ are topologically trivial. Then the restrictions $P\vert_{U_i\times S^2}$ are $\SU{2}$-equivariant bundles which are trivial over $U_i$. Therefore, by the previous paragraph,
\begin{align}
 P\vert_{U_i\times S^2} \cong U_i \times P_{\lambda_i},
\end{align}
where the homomorphisms $\lambda_i\colon\U{1}\to\curlG$ may be different for different open sets $U_i$. However, by looking at non-empty overlaps $U_{ij} = U_i\cap U_j$, one finds,
\begin{align}
 U_{ij} \times P_{\lambda_i} \cong P\vert_{U_{ij}\times S^2} \cong U_{ij} \times P_{\lambda_j}. \label{eq:P_isom}
\end{align} 
This shows that $P_{\lambda_i} \cong P_{\lambda_j}$ and hence $\lambda_i$ and $\lambda_j$ must lie in the same conjugacy class. If $\Sigma$ is connected, which we have assumed as part of the definition of a Riemann surface, we can therefore choose a single $\lambda\colon\U{1}\to\curlG$ such that
\begin{align}
 P\vert_{U_i\times S^2} \cong U_i \times P_{\lambda}.
\end{align}

The isomorphisms in \eqref{eq:P_isom} therefore give rise to an automorphism of $U_{ij} \times P_{\lambda}$ which is determined by a transition function $h_{ij}\colon U_{ij} \to \curlG$ such that
\begin{align}
 \lambda = h_{ij}^{-1} \lambda \, h_{ij},
\end{align}
i.e.~$h_{ij}$ takes values in $\mathcal{C}_{\curlG}(\lambda(\U{1}))$, the centralizer of $\lambda(\U{1})$ in $\curlG$. Furthermore, on triple overlaps $U_i\cap U_j\cap U_k\ne\emptyset$ the \v{C}ech cocycle condition holds,
\begin{align}
 &h_{ik} = h_{ij}h_{jk}.
\end{align}
Thus the $h_{ij}$ define a principal bundle $P_{\Sigma}$ over $\Sigma$ with structure group 
\begin{align}
 \curlH = \mathcal{C}_{\curlG}(\lambda(\U{1})). \label{eq:2d_group}
\end{align}
This centralizer, $\curlH$, is the residual gauge group after dimensional reduction.

Before giving the general form of an $\SU{2}$-invariant connection on $P$, we note that the homomorphism $\lambda$ is determined by a unique $\Lambda\in\mfk{g}$, the Lie algebra of $\curlG$, which is defined as follows: Introduce the Pauli matrices
\begin{align}
 \sigma_1 = \left( \begin{array}{cc} 0 & 1 \\ 1 & 0 \end{array}\right), \quad
 \sigma_2 = \left( \begin{array}{cc} 0 & -\im \\ \im & 0 \end{array}\right), \quad
 \sigma_3 = \left( \begin{array}{cc} 1 & 0 \\ 0 & -1 \end{array}\right).
\end{align}
Then $E_a = -\ihalf\sigma_a$, for $i=1,2,3$, form a basis of the Lie algebra $\su{2}$ and $E_3$ generates the subgroup $\U{1}$ which is the isotropy group at $[\id] \in \SU{2}/\U{1}$. Then, for a matrix $\Lambda\in\mfk{g}$,
\begin{align}
 \lambda\!\left(\e^{E_3 t}\right) = \e^{\Lambda t},
\end{align}
i.e.~$\Lambda$ generates a $\U{1}$ subgroup in $\curlG$ which is the image of $\lambda$. From $\exp(4\pi E_3)=1_2$ it follows that $\Lambda$ must satisfy
\begin{align}
 \e^{4\pi\Lambda} = \id_{\curlG}. \label{eq:4_pi_constr}
\end{align}

Now let $\omega$ be an $\SU{2}$-invariant connection on the equivariant bundle $P$ over $M$. Over the open set $U_i\subset\Sigma$, we can pull $\omega$ back to an $\SU{2}$-invariant connection on $U_i\times P_{\lambda}$, and this connection corresponds to a gauge potential $\curlA_i$ on $U_i\times S^2$ which is given by
\begin{align}
 &\curlA_{i,z} = A_{i,z}(z, \bar z), \label{eq:inv_A_1}\\
 &\curlA_{i,\bar z} = A_{i,\bar z}(z, \bar z), \label{eq:inv_A_2}\\
 &\curlA_{i,y} = \frac{1}{1+y\bar y}\left( -\im\Lambda \bar y - \Phi_i(z, \bar z) \right), \label{eq:inv_A_3}\\ 
 &\curlA_{i,\bar y} = \frac{1}{1+y\bar y}\left( \im\Lambda y + \Phi_i(z, \bar z)^{\dgr} \right), \label{eq:inv_A_4}
\end{align}
subject to the constraints
\begin{align}
  &[\Lambda, A_{i,z}] = [\Lambda, A_{i,\bar z}] = 0, \label{eq:constr_i_1} \\
  &[\Lambda, \Phi_i] = -\im\Phi_i, \quad [\Lambda, \Phi_i^{\dgr}] = \im\Phi_i^{\dgr}.\label{eq:constr_i_2}
\end{align}
The above formulae are a special case of the results derived in \cite{ForgacsManton, Harnad, KapeZoup}, but also compare \cite{GP_invariant, Popov_quiver, DolanSzabo, MantonSakai}. Note that $A_{i,z}$, $A_{i,\bar z}$ and $\Phi_i$ take values in $\mfk{g}^*$, the complexification of the Lie algebra $\mfk{g}$, which is merely a consequence of our choice to express $\curlA_i$ in terms of complex coordinates.  

On non-empty overlaps $U_{ij}$ one finds the relations,
\begin{align}
 &A_j = h_{ij}^{-1} A_i \, h_{ij} + h_{ij}^{-1} \d h_{ij}, \label{eq:A_trafo}\\
 &\Phi_j = h_{ij}^{-1} \Phi_i \, h_{ij}, \label{eq:Phi_trafo}
\end{align}
where the $h_{ij}\colon U_{ij} \to \curlH$ denote the transition functions of $P_{\Sigma}$ defined above, and $A_i=A_{i,z} \d{z} + A_{i,\bar z}\d{\bar z}$ and analogously for $A_j$. Therefore the collection of the local gauge potentials $A_i$ defines a connection on $P_\Sigma$. Note that the constraints \eqref{eq:constr_i_1} imply that the $A_i$ take their values in $\mfk{h}$, the Lie algebra of $\curlH$, which is consistent with $P_\Sigma$ having structure group $\curlH$.
In the same vein, the $\Phi_i$ define a section of the vector bundle $E_\Sigma$ which is associated to $P_\Sigma$ by the adjoint representation of $\curlH$ on $\mfk{g}^*$. In symbols,
\begin{align}
 E_\Sigma = P_\Sigma \times_{\text{ad}} \mfk{g}^*.
\end{align}

To conclude this section, we remark that the inverse operations of restriction and induction work for principal bundles over product spaces in precisely the same way as they do in the vector bundle case, which has been looked at in \cite{Alvarez1, Alvarez2, Popov_quiver, DolanSzabo}. Starting with the $\SU{2}$-equivariant bundle $P$ over $M$, we can define its restriction to $\Sigma\times[\id]$ which we denote as $P\vert_{\Sigma\times[\id]}$. This is a $\U{1}$-equivariant bundle with structure group $\curlG$, where $\U{1}$ acts trivially on the base and its action on the fibre is defined by the homomorphism $\lambda\colon\U{1}\to\curlG$ associated with $P$ in the same way as above. It can be shown that
\begin{align}
 P\vert_{\Sigma\times[\id]} \cong P_{\Sigma}. \label{eq:bundle_restriction}
\end{align}
The inverse operation is given by the formula
\begin{align}
 P = \SU{2} \times_{\lambda} P\vert_{\Sigma\times[\id]}.
\end{align}
However, our construction of the bundle $P_{\Sigma}$ by analyzing the restrictions of $P$ to patches $U_i\times\Sigma$ rather than using restriction and induction has clarified that
\begin{enumerate}
 \item the structure group of $P_{\Sigma}$ can be reduced to $\mathcal{C}_{\curlG}(\lambda(\U{1}))$ and the equivariant connection on $P$ naturally leads to a connection on $P_{\Sigma}$,
 \item there is an associated vector bundle $E_\Sigma$ of which the Higgs field $\Phi$ is a section.
\end{enumerate}
Furthermore, the analysis carried out in this section should generalize to the situation studied in \cite{Bradlow_look}, where $M=\Sigma\times S^2$ is replaced with a flat fibration
\begin{align}
 S^2 \hookrightarrow M \to \Sigma.
\end{align}

\section{Two-block reduction with Yang--Mills gauge group $\SU{N}/\ZZ_N$} \label{sec:reduction} 

To make further progress, we need to solve explicitly the constraints \eqref{eq:constr_i_1}, \eqref{eq:constr_i_2}, and in order to do so we have to make choices for $\Lambda$ and the Yang--Mills gauge group $\curlG$. For the rest of the paper let $\curlG = \SU{N}/\ZZ_N$, which has the Lie algebra $\mfk{g} = \su{N}$. Note that at the level of pure Yang--Mills theory this is locally indistinguishable from the case where the gauge group is $\SU{N}$ since the centre $\ZZ_N$ acts trivially on the gauge potential $\curlA$. However, we will see that we are forced to take $\curlG = \SU{N}/\ZZ_N$ by the geometry of $S^2$.

Now, since $\Lambda\in\su{N}$, it must be an anti-hermitian and traceless $(N\!\times\!N)$-matrix. By conjugating $\Lambda$ with a suitable $\SU{N}$-matrix, we can make $\Lambda$ diagonal and hence we choose
\begin{align}
\Lambda = \im \left(\begin{array}{cc} \al\,1_m & 0 \\ 0 & \be\,1_n \end{array}\right), 
\end{align}
where $N = m+n$ and $\al$, $\be$ are real constants. To allow for non-trivial solutions of the constraint \eqref{eq:constr_i_2}, it is necessary to require $\al - \be = \pm 1$. Restricting attention to $\al - \be = 1$ and using the tracelessness of $\Lambda$, we find 
\begin{align}
 \al = \frac{n}{N}, \quad \be = -\frac{m}{N}. 
\end{align}
We check that this is consistent with \eqref{eq:4_pi_constr},
\begin{align}
 \e^{4\pi\Lambda} = \left(\begin{array}{cc} \e^{4\pi\im \frac{n}{N}}\,1_m & 0 \\ 0 & \e^{-4\pi\im \frac{m}{N}}\,1_n \end{array}\right)
 = \e^{4\pi\im \frac{n}{N}} \left(\begin{array}{cc} 1_m & 0 \\ 0 & 1_n \end{array}\right)
 = \id \in \SU{N}/\ZZ_N,
\end{align}
which clarifies our choice of the Yang--Mills gauge group. Note that for special values of $N$, $m$, $n$ one may be able to choose a bigger gauge group, i.e.~$\SU{N}$ modulo only a subgroup of $\ZZ_N$. For example, if $N$ is even, it is clearly sufficient to mod out by $\ZZ_{N/2}$, and in the case $m=n$, $N=2m$, it is consistent to work with the gauge group $\SU{N}$, as was done in \cite{MantonSakai}. The special case $N=2$, $m=n=1$, gives the traditional reduction of $\SU{2}$-instantons to abelian vortices, which was first discussed in \cite{Witten}. 

The constraints \eqref{eq:constr_i_1}, \eqref{eq:constr_i_2} are now solved by
\begin{align}
 \Phi = \left(\begin{array}{cc} 0 & 0 \\ \phi & 0 \end{array}\right), \quad
 \Phi^{\dgr} = \left(\begin{array}{cc} 0 & \phi^{\dgr} \\ 0 & 0 \end{array}\right), \label{eq:little_higgs} 
\end{align}
and 
\begin{align}
 A = \left(\begin{array}{cc} a & 0 \\ 0 & b \end{array}\right), \label{eq:little_ab}
\end{align}
where $\phi$ is an $(n\!\times\!m)$-matrix-valued field on $\Sigma$ and we have introduced the gauge potentials $a = a_z \d{z} + a_{\bar z}\d{\bar z}$ and $b = b_z \d{z} + b_{\bar z}\d{\bar z}$ on $\Sigma$. The index $i\in I$ on $A$ and $\Phi$ has been omitted since, as a consequence of \eqref{eq:A_trafo}, \eqref{eq:Phi_trafo}, the constraints \eqref{eq:constr_i_1}, \eqref{eq:constr_i_2} are globally meaningful.

The fields $a$, $b$, $\phi$ are the content of the theory on $\Sigma$ arising as the symmetry reduction of Yang--Mills theory on $M$. The gauge group of the theory on $\Sigma$ coincides with the structure group of $P_\Sigma$ from \eqref{eq:2d_group}, and with our choice of $\Lambda$ is 
\begin{align}
 \curlH = \mathrm{S}\!\left(\U{m}\times\U{n}\right)\!/\ZZ_N, \label{eq:2d_H}
\end{align}
where the leading $\mathrm{S}$ indicates that the overall determinant is one. Modding out by $\ZZ_N$ is meaningful since $\ZZ_N$ is contained in $\mathrm{S}\!\left(\U{m}\times\U{n}\right)$ as a normal subgroup. Because of the structure of $\curlH$, the transition functions $h_{ij}$ of $P_{\Sigma}$ can be written as
\begin{align}
 h_{ij}(z,\bar z) = \left(\begin{array}{cc} h^m_{ij}(z,\bar z) & 0 \\ 0 & h^n_{ij}(z,\bar z) \end{array}\right) \e^\frac{2\pi\im k}{N}, \label{eq:2d_trans}
\end{align}
where $h^m_{ij}\in\U{m}$ and $h^n_{ij}\in\U{n}$ such that $\det{h^m_{ij}}\det{h^n_{ij}} = 1$, and $k$ is an integer between $0$ and $N-1$. The transformation law for the section $\phi$ on $U_{ij}$ can then be read off from \eqref{eq:Phi_trafo},
\begin{align}
 \phi_j = \left(h^n_{ij}\right)^{-1}\phi_i \,h^m_{ij}, \label{eq:phi_trans}
\end{align}
where $\phi_i$ and $\phi_j$ are local expressions for $\phi$ on the open sets $U_i$ and $U_j$ respectively. We thus obtain a more refined picture of $\Phi$ and $E_\Sigma$: There exists a vector bundle $E_{mn}$ over $\Sigma$ of rank $mn$. The structure group of $E_{mn}$ is also $\mathrm{S}(\U{m}\times\U{n})/\ZZ_N$, and its fibres transform according to the law \eqref{eq:phi_trans}. The collection of the $\phi_i$ then comprise a section $\phi$ of $E_{mn}$, and we shall refer to $\phi$ as the non-abelian Higgs field. The gauge potential $A$ from \eqref{eq:little_ab} defines a covariant derivative on $E_{mn}$ by virtue of
\begin{align}
D\Phi = \d\Phi + [A, \Phi] = \left(\begin{array}{cc} 0 & 0 \\ D\phi & 0 \end{array}\right),
\end{align}
with $D\phi = \d\phi +b\phi - \phi a$. From this we can calculate the curvature of the bundle $E_{mn}$, which we denote as $f$. The curvature acts on sections as an endomorphism in the following way,
\begin{align}
 f\phi = f^b\phi - \phi f^a,
\end{align}  
where $f^a = \d a + a\w a$ and $f^b = \d b + b\w b$. Then a straightforward calculation shows that 
\begin{align}
 \tr{f} = m\,\tr{f^b} - n\,\tr{f^a},
\end{align}
where on the left hand side the trace is taken in the space of endomorphisms of $(n\!\times\! m)$-matrices. Thus, the first Chern number of $E_{mn}$ is
\begin{align}
 c_1(E_{mn}) &= \frac{\im}{2\pi} \int_{\Sigma} \tr{f_{z\bar z}} \d{z}\w\d{\bar z} \\
 	     &= \frac{\im}{2\pi} \int_{\Sigma} \left(m\,\tr{f_{z \bar z}^b} - n\,\tr{f_{z \bar z}^a}\right) \d{z}\w\d{\bar z}.
\end{align}
Since the non-abelian Higgs field $\phi$ is a section of $E_{mn}$, we will take $c_1(E_{mn})$ as a generalization of the vortex number in the abelian Higgs model. This will be motivated more in the next section, where we relate $c_1(E_{mn})$ to the instanton number of the Yang--Mills theory on $\Sigma\times S^2$.

The above expression for $c_1(E_{mn})$ can be simplified: Note that the gauge potential $A$ is traceless, and therefore also $\tr{f^a} + \tr{f^b} = 0$. Hence,
\begin{align}
 c_1(E_{mn}) = \frac{\im N}{2\pi} \int_{\Sigma} \tr{f_{z \bar z}^b} \d{z}\w\d{\bar z} 
	     = -\frac{\im N}{2\pi} \int_{\Sigma} \tr{f_{z \bar z}^a} \d{z}\w\d{\bar z}.
\end{align}
It follows from the general theory of complex vector bundles that $c_1(E_{mn})$ is an integer. Therefore,
\begin{align}
 \frac{\im}{2\pi} \int_{\Sigma} \tr{f_{z \bar z}^a} \d{z}\w\d{\bar z} \:\in\frac{1}{N}\ZZ, \quad
 \frac{\im}{2\pi} \int_{\Sigma} \tr{f_{z \bar z}^b} \d{z}\w\d{\bar z} \:\in\frac{1}{N}\ZZ. \label{eq:fa_fb_NZ}
\end{align}
Since these expressions need not be integral, this shows that in general there are no vector bundles with gauge potentials $a$ and $b$. This is in agreement with the general form of the transition function \eqref{eq:2d_trans}: The entries $h^m_{ij}$ and $h^n_{ij}$ need not satisfy the \v{C}ech cocycle conditions in $\U{m}$ or $\U{n}$ respectively, but only up to an element of $\ZZ_N$. As a consequence, unlike in \cite{GP_invariant, Alvarez1, Popov_quiver, DolanSzabo}, $E_{mn}$ cannot be thought of as the bundle of homomorphisms between two distinct vector bundles over $\Sigma$. This can be traced back to the fact that since we consider Yang--Mills theory with gauge group $\SU{N}/\ZZ_N$, there is no vector bundle of rank $N$ which is naturally associated with this Yang--Mills theory on $\Sigma\times S^2$.

\section{Actions, energy and vortex equations} \label{sec:actions}

We denote the field strength of Yang--Mills theory as $\curlF = \d\curlA + \curlA\w\curlA$. The action of Yang--Mills theory on $\Sigma\times S^2$ in terms of the complex coordinates $z$, $y$ is given by 
\begin{align}
  S_{\textit{YM}} = \int_{\Sigma\times S^2} \mathrm{tr}\bigg( \frac{4}{\sigma^2}\curlF_{z\bar z}^2 &- 
	\frac{(1+y\bar y)^2}{\sigma} \left(\curlF_{zy}\curlF_{\bar z \bar y}+ \curlF_{z\bar y}\curlF_{\bar z y}\right) \nonumber\\
	&+ \frac{(1+y\bar y)^4}{16}\curlF_{y\bar y}^2\bigg)  \, \dvol_{\Sigma}\,\dvol_{S^2}. 
\end{align}
One should bear in mind that $S_{\textit{YM}}$ is non-negative since the components of $\curlF$ in real directions are anti-hermitian for unitary gauge groups such as $\SU{N}/\ZZ_N$. Substituting in the symmetric ansatz for $\curlA$ from \eqref{eq:inv_A_1}-\eqref{eq:inv_A_4} we obtain the reduced action on $\Sigma$,
\begin{align}
 S_{\Sigma} = 8\pi \int_{\Sigma} \mathrm{tr}\bigg( \frac{4}{\sigma^2}F_{z\bar z}^2 &+
	  \frac{1}{\sigma}\left(D_z\Phi D_{\bar z}\Phi^{\dgr} + D_{\bar z}\Phi D_z \Phi^{\dgr} \right) \nonumber\\
	  &+ \frac{1}{16}\left(2\im\Lambda - [\Phi,\Phi^{\dgr}] \right)^2 \bigg) \, \sigma\,\d{x^1}\w\d{x^2}, \label{eq:inv_S_YM}
\end{align}
where $F = \d A + A\w A$ and we have performed the integral over $S^2$,
\begin{align}
 \int_{S^2} \frac{8}{(1+y\bar y)^2} \d{x^3}\w\d{x^4} = 8\pi.
\end{align}
It is sensible to identify $S_{\Sigma}$ with the potential energy $E$ of a field theory on $\Sigma$ since it is a static action. We use the convention $E = S_{\Sigma}/ 16\pi$, where we have divided by the area of $S^2$ and introduced a factor of $\half$. Henceforth we shall assume $m\ge n$. Then we can express the energy $E$ in terms of the unconstrained fields $a$, $b$, $\phi$ on $\Sigma$, 
\begin{align}
  E = \half \int_{\Sigma} \bigg( &\frac{4}{\sigma^2} \left(\tr{f_{z\bar z}^a f_{z\bar z}^a} 
	    + \tr{f_{z\bar z}^bf_{z\bar z}^b} \right) \nonumber\\
	    &+ \frac{1}{\sigma}\left(\tr{D_z\phi D_{\bar z}\phi^{\dgr}} + \tr{D_{\bar z}\phi D_z\phi^{\dgr}} \right) \nonumber\\
	    &+ \frac{1}{8}\frac{n(m-n)}{N}  + \frac{1}{8} \tr{ 1_n - \phi\phi^{\dgr}}^2 \bigg) \, \sigma\,\d{x^1}\w\d{x^2},
	\label{eq:2d_enery}
\end{align}
where the constant term arises because
\begin{align}
 \tr{2\im\Lambda - [\Phi,\Phi^{\dgr}] }^2
 &= \tr{-2\frac{n}{N}1_m + \phi^{\dgr}\phi}^2 + \tr{2\frac{m}{N}1_n - \phi\phi^{\dgr}}^2 \\
 &= 2\frac{n(m-n)}{N}  + 2\, \tr{ 1_n - \phi\phi^{\dgr}}^2.
\end{align}
From the above expression for $E$ we can immediately read off the lower bound
\begin{align}
 E \ge \frac{A_{\Sigma}}{16N} n(m-n), \label{eq:bd_apriori}
\end{align}
which we shall call the \textit{Bimonte--Lozano} bound since a similar bound was derived in \cite{BimonteLozano} in a purely two-dimensional context. This lower bound for $E$ provides good motivation for putting the theory on compact $\Sigma$, as here and in \cite{BimonteLozano}. If $\Sigma$ had infinite area, then for $m\ne n$ the energy of the theory on $\Sigma$ would always be infinite. The action $S_{\textit{YM}}$ of the corresponding Yang--Mills theory on $\Sigma\times S^2$ would also be infinite, and hence instantons with $\SU{2}$-symmetry and with this choice of $\Lambda$ do not contribute to the partition function. From a two-dimensional point of view this infinity can of course be cured by subtracting a constant from $E$, but from a four-dimensional point of view this is unnatural. 

It is clear from \eqref{eq:bd_apriori} that $E=0$ cannot be achieved for $m\ne n$. This is not in disagreement with the fact that Yang--Mills theory on $\Sigma\times S^2$ always admits the vacuum solution $\curlF=0$ since, for $m\ne n$, this solution does not lie in the sector of $\SU{2}$-invariant solutions we are studying. We can make this statement more precise: By looking at \eqref{eq:inv_S_YM} we see that if the vacuum of Yang--Mills theory is an $\SU{2}$-invariant field configuration, the fields $A$ and $\Phi$ on $\Sigma$ satisfy
\begin{align}
  &F_{z \bar z}=0, \\
  &D_{z}\Phi = D_{\bar z}\Phi = 0, \\
  &[\Phi,\Phi^{\dgr}] = 2\im\Lambda.
\end{align}
The last condition, together with \eqref{eq:constr_i_2}, implies that $\Lambda$, $\Phi$, $\Phi^{\dgr}$ form an $\sli{2}$-representation\footnote{It is an $\sli{2}$-representation rather than an $\su{2}$-representation, as one might have expected, because $\Phi$ takes values in the complexified Lie algebra $\mfk{g}^*$.}. One can easily check algebraically that this can only be the case when $m=n$, in agreement with \eqref{eq:bd_apriori}.

We now turn to the self-duality equations of Yang--Mills theory on $\Sigma\times S^2$, which in our conventions read
\begin{align}
 \frac{8}{(1+y\bar y)^2}&\curlF_{z\bar z} = \sigma \curlF_{y \bar y}, \label{eq:SD_1}\\
 &\curlF_{z\bar y} = 0, \label{eq:SD_2}\\
 &\curlF_{\bar{z} y} = 0. \label{eq:SD_3}
\end{align}
Substituting in \eqref{eq:inv_A_1}-\eqref{eq:inv_A_4}, these reduce to
\begin{align}
 &F_{z\bar z} = \frac{\sigma}{8}\left(2\im\Lambda - [\Phi,\Phi^{\dgr}] \right), \label{eq:sym_SD_1}\\
 &D_{z}\Phi^{\dgr} = 0, \label{eq:sym_SD_2}\\
 &D_{\bar z}\Phi = 0, \label{eq:sym_SD_3}
\end{align}
 and these equations must of course be supplemented with the constraints \eqref{eq:constr_i_1}, \eqref{eq:constr_i_2}. Solving the constraints, as we have done in the previous section, we obtain the Bogomolny-type vortex equations
\begin{align}
 &f^a_{z\bar z} = \frac{\sigma}{8} \left(-\frac{2n}{N} 1_m + \phi^{\dgr} \phi\right), \label{eq:Bog_1}\\
 &f^b_{z\bar z} = \frac{\sigma}{8} \left( \frac{2m}{N} 1_n - \phi \phi^{\dgr}\right), \label{eq:Bog_2}\\
 &D_{\bar z}\phi = 0 \label{eq:Bog_3},
\end{align}
where $f^a$, $f^b$ and $D$ are as in the previous section. 

When the self-duality equations \eqref{eq:SD_1}-\eqref{eq:SD_3} are satisfied, the Yang--Mills action is $S_{\textit{YM}} = -8\pi^2 c_2$ with
\begin{align}
 c_2 &= \frac{1}{8\pi^2} \int_{\Sigma\times S^2} \tr{\curlF\w\curlF}
\end{align}
the instanton number. Note that $c_2$ must be negative or zero since we have already established that $S_{\textit{YM}}$ is non-negative. We should stress that in the case we are interested in, this formula cannot always be expected to yield an integral value for $c_2$. This is because $\curlF$ is not the curvature of a vector bundle but only of the principal bundle $P$ with structure group $\SU{N}/\ZZ_N$. Substituting again \eqref{eq:inv_A_1}-\eqref{eq:inv_A_4} into the expression for $c_2$, one obtains
\begin{align}
  c_2 &= \frac{\im}{\pi N} \int_{\Sigma} \left( n\, \tr{f_{z\bar z}^a} - m \, \tr{f_{z\bar z}^b} \right)
	 \d{z}\w\d{\bar z}. \label{eq:c2_traces} \\
      &= -\frac{2}{N} c_1(E_{mn}) \label{eq:c2_c1},
\end{align}
so $c_2$ is $\frac{1}{N}$ times an integer.
By a Bogomolny-type argument one can show that $E$ reduces to 
\begin{align}
 E = \frac{\pi}{N} c_1(E_{mn}), \label{eq:energy_c1}
\end{align}
when the Bogomolny equations \eqref{eq:Bog_1}-\eqref{eq:Bog_3} are satisfied. Note that in deriving \eqref{eq:energy_c1} from the general expression for $E$ we have dropped a boundary term, which is certainly valid since $\Sigma$ is assumed a closed, compact surface. 

Equations \eqref{eq:c2_c1} and \eqref{eq:energy_c1} strengthen our interpretation of $c_1$ as the non-abelian vortex number. By \eqref{eq:energy_c1} we also assign the mass $\pi/N$ to every vortex. In the case $N=2$, this leads to a mass of $\pi/2$ for the single abelian vortex. We would have obtained the more usual answer $\pi$ (cf.~\cite{MantonSut}) had we chosen a different convention for the energy, namely $E = S_{\Sigma}/8\pi$.

\section{Energy bounds for solutions of the Bogomolny equations} \label{sec:bounds}

For solutions which satisfy the Bogomolny equations \eqref{eq:Bog_1}-\eqref{eq:Bog_3}, we can obtain a sharper lower bound for $E$ than the one in \eqref{eq:bd_apriori} by replacing $f^a$ and $f^b$ in $E$ by the right hand sides of \eqref{eq:Bog_1} and \eqref{eq:Bog_2} respectively. We find
\begin{align}
  E \ge \frac{A_{\Sigma}}{8N} n(m-n). \label{eq:bd_sd}
\end{align}
In section \ref{sec:max_abelian} we will give solutions to the Bogomolny equations which saturate this bound, but it is clear that these solutions cannot saturate \eqref{eq:bd_apriori} unless $m=n$. Whether for $m\ne n$ the theory defined by $E$ has solutions which do not satisfy the Bogomolny equations but saturate \eqref{eq:bd_apriori}, we do not know.

An upper bound for $E$ can also be derived when the Bogomolny equations are satisfied. From \eqref{eq:energy_c1} we get 
\begin{align}
 E = -\frac{1}{N} \int_{\Sigma} \left( n\, \tr{f_{z\bar z}^a} - m \, \tr{f_{z\bar z}^b} \right)
 \d{x^1}\w\d{x^2},
\end{align}
and again substituing in \eqref{eq:Bog_1}, \eqref{eq:Bog_2} for $f^a$ and $f^b$,
\begin{align}
 E &= \frac{1}{8N} \int_{\Sigma} \left( 2nm - N\,\tr{\phi^{\dgr} \phi} \right) 
 \sigma\d{x^1}\w\d{x^2} \\
   &\le \frac{nm}{4N} A_{\Sigma}. \label{eq:bd_upper}
\end{align}
This generalizes the well-known Bradlow bound \cite{Bradlow, GP_exist} in the abelian Higgs model.

Finally, combining these energy bounds for solutions of the Bogomolny equations, we obtain
\begin{align}
 \frac{c_1(E_{mn})}{mn} \le \frac{A_{\Sigma}}{4\pi} \le \frac{2c_1(E_{mn})}{n(m-n)},
\end{align}
which relates the topology of a solution, captured by $c_1(E_{mn})$, to the geometry of $\Sigma$, captured by its area.

\section{Solutions of maximally abelian type} \label{sec:max_abelian}

We now construct special solutions to the coupled vortex equations \eqref{eq:Bog_1}-\eqref{eq:Bog_3} by embedding as many abelian Higgs fields as possible into the non-abelian Higgs field $\phi$. Hence we refer to this special class of solutions as being \textit{maximally abelian}. 

We start with the following ansatz for the gauge potential $A$,
\begin{align}
  &A = \left(\begin{array}{ccccccc}
 		a_1 &            &        &               &        &           &      \\
		       & \ddots &        &               &        &           &       \\
		       &            & a_n &               &        &           &       \\
		       &            &        & \begin{xy} (0,0)*+{\al}*\frm{-} \end{xy}  &        &            &       \\
		       &            &        &              &  b_1 &            &       \\
		       &            &        &              &        & \ddots &        \\
		       &            &        &              &        &            & b_n
		       \end{array}\right),
\end{align}
where $a_1, \dots, a_n, b_1, \dots, b_n \in \u{1}$, $\al\in\u{m-n}$, and $\sum_{i=1}^n (a_i+b_i) + \tr{\al} = 0$. (Note that the subscript $i$ no longer refers to sets $U_i$ in an open covering of $\Sigma$ as in sections \ref{sec:ansatz} and \ref{sec:reduction}.) This corresponds to assuming that the structure group of $E_{mn}$ can be reduced to 
\begin{align}
 \mathrm{S}(\U{1}^n\times\U{m-n}\times\U{1}^n)/\ZZ_N.
\end{align}
The Higgs field $\phi$ is then chosen to be as diagonal as possible, i.e.
\begin{align}
 \phi = \left(\begin{array}{ccc|c}
 		\phi_1 &            &            &         \\
		           & \ddots &            &   0    \\
		           &            & \phi_n &        
		       \end{array}\right), 
\end{align}
where the entries $\phi_i$, $i=1,\dots,n$, can be interpreted as sections of appropriate complex line bundles over $\Sigma$, and the zero entries should be regarded as zero sections. Inserting our ansatz into \eqref{eq:Bog_1}-\eqref{eq:Bog_3} yields
\begin{align}
 &f^{a_i}_{z\bar z} = \frac{\sigma}{8}\left(-\frac{2n}{N} + \modsq{\phi_i}\right), \label{eq:ab_Bog_1} \\
 &f^{b_i}_{z\bar z} = \frac{\sigma}{8}\left( \frac{2m}{N} - \modsq{\phi_i}{}\right), \label{eq:ab_Bog_2} \\
 &f^{\al}_{z\bar z} = -\sigma \frac{n}{4N} 1_{m-n}, \label{eq:const_flux} \\
 &\pd_{\bar z}\phi_i + (b_{i\bar z} - a_{i \bar z})\phi_i = 0,
\end{align}
where $f^{a_i} = \d a_i$, $f^{b_i} = \d b_i$, for $i=1,\dots,n$, and $f^{\al} = \d\al + \al\w\al$. It is convenient to arrange these equations into two sets by adding and subtracting the equations for $f^{a_i}$ and $f^{b_i}$, which is allowed since these are abelian field strengths and globally well-defined. Thus, the above set of equations is equivalent to
\begin{align}
 &f^{b_i}_{z\bar z} - f^{a_i}_{z\bar z} = \frac{\sigma}{4}\left( 1 - \modsq{\phi_i}\right), \label{eq:ab_Higgs_1}\\
 &\pd_{\bar z}\phi_i + (b_{i\bar z} - a_{i \bar z})\phi_i = 0, \label{eq:ab_Higgs_2}
\end{align}
and
\begin{align}
 f^{b_i}_{z\bar z} + f^{a_i}_{z\bar z} &= \frac{\sigma}{4} \frac{m-n}{N}, \label{eq:f_sum}\\
 f^{\al}_{z\bar z} &= -\sigma \frac{n}{4N} 1_{m-n}. \label{eq:f_non_ab}
\end{align}
Equations \eqref{eq:ab_Higgs_1}, \eqref{eq:ab_Higgs_2} are readily identified as $n$ independent pairs of Bogomolny equations of the abelian Higgs model with Higgs fields $\phi_i$, $i=1,\dots,n$. These equations can be solved on the Riemann surface $\Sigma$ provided the Bradlow inequality holds,
\begin{align}
 r_i = \frac{\im}{2\pi} \int_{\Sigma} (f^{b_i}_{z\bar z}-f^{a_i}_{z\bar z}) \, \d{z}\w\d{\bar z} \le \frac{1}{4\pi}A_{\Sigma}, \label{eq:Bradlow}
\end{align}
as was shown in \cite{Bradlow, GP_exist}. We have $r_i\in\ZZ$ since $r_i$ is the vortex number of the abelian Higgs field $\phi_i$ and it can also be identified as the first Chern number of a line subbundle of $E_{mn}$.

Equations \eqref{eq:f_sum} and \eqref{eq:f_non_ab} define magnetic fields on $\Sigma$ with constant flux densities. The solution theory that was established in \cite{Bradlow, GP_exist} can also be applied to \eqref{eq:f_sum} and \eqref{eq:f_non_ab}, and it follows that a necessary and sufficient condition for the existence of solutions is obtained by integrating \eqref{eq:f_sum} and \eqref{eq:f_non_ab},
\begin{align}
 &r_i' = \frac{\im}{2\pi} \int_{\Sigma} (f^{b_i}_{z\bar z}+f^{a_i}_{z\bar z}) \, \d{z}\w\d{\bar z} 
       = A_{\Sigma} \frac{(m-n)}{4\pi N} \label{eq:r_prime}, \\
 &r =  \frac{\im}{2\pi} \int_{\Sigma} \tr{f^{\al}_{z\bar z}} \d{z}\w\d{\bar z} 
    = - A_{\Sigma} \frac{n(m-n)}{4\pi N} \label{eq:r_wind}.
\end{align}
The $r_i'$ and $r$ take values in $\frac{1}{N}\ZZ$ since they can be regarded as first Chern numbers of principal bundles over $\Sigma$ with structure groups $\U{1}/\ZZ_N$ and $\U{m-n}/\ZZ_N$ respectively. This puts constraints on the area $A_{\Sigma}$. Another constraint is obtained by the integrality of $c_1(E_{mn})$ and the formula
\begin{align}
 c_1(E_{mn}) &= \frac{\im}{2\pi} \int_{\Sigma} \left( m \sum_{i=1}^n f^{b_i}_{z \bar z} 
 					- n \sum_{i=1}^n f^{a_i}_{z \bar z} 
					- n\,\tr{f^{\al}_{z \bar z}}
					\right) \d{z}\w\d{\bar z}, \\ 
	&= \frac{N}{2} \sum_{i=1}^n r_i + \frac{n(m-n)}{8\pi}A_{\Sigma}. \label{eq:c1_vortices}
\end{align}
This is easily converted into an expression for the energy,
\begin{align}
 E = \frac{\pi}{2} \sum_{i=1}^n r_i + \frac{n(m-n)}{8N}A_{\Sigma}.
\end{align}
Solutions with $r_i=0$ for all $i=1,\dots,n$ saturate the lower bound \eqref{eq:bd_sd} for $E$, and $r_i=0$ is achieved by Higgs fields $\phi_i$ with $\md{\phi_i}=1$. The upper bound \eqref{eq:bd_upper} for $E$ can only be saturated if, in addition to \eqref{eq:r_prime} and \eqref{eq:r_wind}, one can satisfy $r_i = A_{\Sigma}/4\pi$. 
Note also that the first term in $E$ is the sum of the masses of the vortices in the various abelian Higgs models corresponding to $i=1,\dots,n$. Rather remarkably this contribution to $E$ is independent of $N$.

\section{Examples: The cases $N=3$ and $N=5$} \label{sec:examples}

Part of the motivation for the present work was drawn from \cite{BimonteLozano}, where an $\SU{2}\times\U{1}$ Yang--Mills--Higgs model was studied. This is the well known electroweak sector of the Standard Model, and in \cite{BimonteLozano} this theory was considered on a flat torus to investigate periodic vortex solutions on the plane $\RR^2$. We now show how this can be understood as invariant Yang--Mills theory on $\Sigma\times S^2$, with $\Sigma$ a torus.

In order to have the right symmetry breaking pattern, we take $N=3$, $m=2$, $n=1$. Then, 
\begin{align}
 \SU{3}/\ZZ_3 \to \textrm{S}(\U{2}\times\U{1})/\ZZ_3 \approx \SU{2}\times\U{1},
\end{align}
where $\approx$ denotes a local isomorphism. A basis of the real Lie algebra $\su{3}$ is given by the anti-hermitian matrices $-\ihalf\lambda_r$, $r=1,\dots,8$, where the $\lambda_r$ are the hermitian Gell-Mann matrices. These are
\begin{align}
 &\lambda_1 = \left(\begin{array}{ccc} 0 & 1 & 0 \\ 1 & 0 & 0 \\ 0 & 0 & 0 \end{array}\right), \quad
  \lambda_2 = \left(\begin{array}{ccc} 0 & -\im & 0 \\ \im & 0 & 0 \\ 0 & 0 & 0 \end{array}\right), \\
 &\lambda_3 = \left(\begin{array}{ccc} 1 & 0 & 0 \\ 0 & -1 & 0 \\ 0 & 0 & 0 \end{array}\right), \quad
  \lambda_8 = \frac{1}{\sqrt{3}}\left(\begin{array}{ccc} 1 & 0 & 0 \\ 0 & 1 & 0 \\ 0 & 0 & -2 \end{array}\right), \\ 
 &\lambda_4 - \im\lambda_5 = \left(\begin{array}{ccc} 0 & 0 & 0 \\ 0 & 0 & 0 \\ 2 & 0 & 0 \end{array}\right), \quad
  \lambda_6 - \im\lambda_7 = \left(\begin{array}{ccc} 0 & 0 & 0 \\ 0 & 0 & 0 \\ 0 & 2 & 0 \end{array}\right). 
\end{align}
Then $\Lambda = \im \lambda_8/\sqrt{3}$, and the following commutation relations hold:
\begin{align}
 &[\lambda_r,\lambda_s] = 2\im\,\epsi_{rst}\lambda_t, \quad 
 [\lambda_8,\lambda_r] = 0, \\
 &[\lambda_8,\lambda_4-\im\lambda_5] = -\sqrt{3}(\lambda_4-\im\lambda_5), \quad
 [\lambda_8,\lambda_6-\im\lambda_7] = -\sqrt{3}(\lambda_6-\im\lambda_7),
\end{align}
for $r,s,t=1,2,3$. From now on, in this section, the lower case Latin indices $r,s,t$ will always run over $1,2,3$.

The gauge potential $A$ from \eqref{eq:little_ab} is written in terms of the $\lambda_r$ and $\lambda_8$ as
\begin{align}
 A = -\ihalf A^r\lambda_r -\ihalf A^8\lambda_8,
\end{align}
i.e.
\begin{align}
 a = -\ihalf A^r\sigma_r - \frac{\im}{2\sqrt{3}} A^8, \quad b = \frac{\im}{\sqrt{3}} A^8, 
\end{align}
where the $\sigma_r$ are the Pauli matrices from section \ref{sec:ansatz}. The Higgs field $\phi$ is a two-component row vector, $\phi = (\phi_1, \phi_2)$, and from \eqref{eq:little_higgs} we see that
\begin{align}
 \Phi = \half \phi_1 (\lambda_4 - \im\lambda_5) + \half \phi_2 (\lambda_6 - \im\lambda_7).
\end{align}
From the covariant derivative of $\phi$,
\begin{align}
 D\phi = \d\phi + \ihalf\sqrt{3}A^8\phi + \ihalf A^r\phi\sigma_r,
\end{align}
we can determine the Weinberg angle $\th_W = 60^{\circ}$, in agreement with the earlier result in \cite{Manton_6d}.

Since $\Sigma$ is a torus, the conformal factor $\sigma$ is a constant, and we set $\sigma=1$. If we also perform the rescaling $\phi = \sqrt{2}\,\tilde{\phi}$, the expression for the energy \eqref{eq:2d_enery} becomes 
\begin{align}
 E = \frac{1}{2} \int_{\Sigma} \bigg( &\frac{1}{4} \left({f_{ij}^r f_{ij}^r} 
	    +  f_{ij}^8f_{ij}^8 \right) 
	    +  D_i\tilde{\phi}(D_i\tilde{\phi})^{\dgr} \nonumber\\
	    &+ \frac{1}{2} \left(\frac{1}{2} - \tilde{\phi}\tilde{\phi}^{\dgr}\right)^2 + \frac{1}{24}\bigg) \, \d{x^1}\w\d{x^2}, \label{eq:BL_energy}
\end{align}
where $i,j=1,2$ denote the real directions on $\Sigma$, and the field strengths are defined as
\begin{align}
 &f_{ij}^r = \pd_i A_j^r - \pd_j A_i^r + \epsi_{rst}A_i^s A_j^t, \\
 &f_{ij}^8 = \pd_i A_j^8 - \pd_j A_i^8.
\end{align}
The rescaling $\phi = \sqrt{2}\,\tilde{\phi}$ was necessary to bring the kinetic term for the Higgs field in \eqref{eq:BL_energy} into canonical form; it is also very natural since $\sqrt{2}$ is the radius of the internal $S^2$ and introducing a length scale is required to ensure that $\tilde{\phi}$ has the right dimensions for a complex scalar in two dimensions (see \cite{DolanSzabo} for details). It is clear from \eqref{eq:BL_energy} that the radius of the internal $S^2$ determines the mass of the Higgs field, $m_H = 1$. The Z-boson mass in these units is also $m_Z = 1$.

We discard the constant term in \eqref{eq:BL_energy} by defining the renormalized energy,
\begin{align}
 E_{\textit{ren}} = E - \frac{1}{48} A_\Sigma.
\end{align}
When the Bogomolny equations are satisfied this yields
\begin{align}
 E_{\textit{ren}} = -\frac{1}{2\sqrt{3}}\int_{\Sigma}f_{12}^8 \,\d{x^1}\w\d{x^2} - \frac{1}{48} A_\Sigma,
\end{align}
where the first term is just \eqref{eq:energy_c1} with the Chern number written explicitly in terms of the abelian flux density $f_{12}^8$. The above expression for $E_{\textit{ren}}$ agrees with the energy bound in \cite{BimonteLozano} in the right units. It can also be shown that our Bogomolny equations \eqref{eq:Bog_1}-\eqref{eq:Bog_3} are the same as in \cite{BimonteLozano}. 

The field configuration in the $N=3$, $m=2$, $n=1$ case, on a generic Riemann surface $\Sigma$, is maximally abelian if the Higgs field can globally be written as $\phi = (\phi_1,0)$. According to the previous section, the entry $\phi_1$ must satisfy the Bogomolny equations of the abelian Higgs model, and the area $A_{\Sigma}$ is subject to various constraints. Let us fix the area to be $A_{\Sigma} = 4\pi k$ with $k\in\NN$. Then the Bradlow bound \eqref{eq:Bradlow} can be saturated. Also, the constraints \eqref{eq:r_prime}, \eqref{eq:r_wind} are solved by setting $3r_1'=-3r=k$, in agreement with $r_1'$, $r$ being fractions with denominator $N=3$. In this situation we obtain for the energy, 
\begin{align}
 E = \frac{\pi}{2}r_1 + \frac{1}{24}A_{\Sigma},
\end{align}
with the vortex number $r_1$ counting the zeros of the abelian Higgs field $\phi_1$. The lower bound $E = A_{\Sigma}/24$ is attained for $r_1=0$, and the upper bound $E = A_{\Sigma}/6$ is attained for $r_1=k$, when the Bradlow bound is saturated. It should be noted that from the integrality of the first Chern number \eqref{eq:c1_vortices} we obtain the condition
\begin{align}
 3r_1 + k \in 2\ZZ.
\end{align}
Hence, $r_1=0$ is possible only if $k$ is even.

We also comment briefly on the case $N=5$, $m=3$, $n=2$, which may also be of interest in the context of unification of the Standard Model gauge groups. In the maximally abelian situation the Higgs field is 
\begin{align}
 \phi = \left(\begin{array}{ccc} \phi_1 & 0 & 0 \\ 0 & \phi_2 & 0 \end{array}\right),
\end{align}
where the vortex numbers corresponding to the entries $\phi_1$ and $\phi_2$ are $r_1$, $r_2$ respectively. As before we choose $A_{\Sigma} = 4\pi k$ so that the Bradlow bound can be saturated. Then \eqref{eq:r_prime}, \eqref{eq:r_wind} are satisfied by
\begin{align}
 5r_1' = 5r_2' = k, \quad -5r = 2k. 
\end{align}
The allowed energy range is 
\begin{align}
 \frac{1}{20} A_{\Sigma} \le E \le \frac{3}{10} A_{\Sigma},
\end{align}
and the lower and upper bounds are attained when $r_1 = r_2 = 0$, and when $r_1 = r_2 = k$ respectively. 
Integrality of the  Chern number \eqref{eq:c1_vortices} leads to the condition 
\begin{align}
 5(r_1+r_2) \in 2\ZZ,
\end{align}
which, contrary to the $N=3$ case, allows $r_1 = r_2 = 0$ for arbitrary $k$.

Finally we stress again that the case $N=2m$, $m=n$, has been studied for general $m$ in \cite{MantonSakai}. We have noted before (see section \ref{sec:reduction}) that in this case it is not necessary to divide the Yang--Mills gauge group by $\ZZ_{2m}$, and one can choose to work with $\SU{2m}$, as in \cite{MantonSakai}. Nonetheless we can apply our analysis, and, most notably, for $m=n$ the energy does not include a term proportional to $A_{\Sigma}$. Hence the lower bound for the energy is zero and is attained by the Yang--Mills vacuum $\curlF = 0$, which is consistent with the discussion following \eqref{eq:bd_apriori}.

\section{Relation to quiver bundles} \label{sec:quivers}

Dimensional reduction of Yang--Mills theory on spaces of the form $\Sigma\times S^2$, where $\Sigma$ is not necessarily a Riemann surface, has received considerable attention in the literature \cite{GP_invariant, Alvarez1, Popov_quiver, Popov_nonab, DolanSzabo}. The focus of previous work in the physics literature, however, has mostly been on Yang--Mills theory with gauge group $\U{N}$. It is then very natural to associate a rank $N$ vector bundle with Yang--Mills theory, and the Yang--Mills gauge potential defines a covariant derivative on this bundle. Invariance under the $\SU{2}$-action is realized by considering $\SU{2}$-equivariant vector bundles, and the reduction to a theory on $\Sigma$ leads to quivers of vector bundles over $\Sigma$, or quiver bundles for short \cite{Alvarez1, Popov_quiver}. 

We can translate our work into the language of quiver bundles by associating a vector bundle $\curlE$ to the $\SU{N}/\ZZ_N$-principal bundle $P$ over $\Sigma\times S^2$. To do so, we have to choose a representation of $\SU{N}/\ZZ_N$, and we take the adjoint representation on $\Endo{\CC^N}$, the space of complex $(N\!\times\!N)$-matrices. This representation is given by
\begin{align}
 \ad{S}M = S\,M\,S^{-1}, \quad S\in\SU{N},\: M \in\Endo{\CC^N}, 
\end{align}
which descends to a representation of $\SU{N}/\ZZ_N$ because $\ad{S\e^\frac{2\pi\im k}{N}} = \ad{S}$ for any integer $k$. The vector bundle $\curlE$ is defined as $\curlE = P \times_{\text{ad}} \Endo{\CC^N}$, and this is naturally an $\SU{2}$-equivariant vector bundle if $P$ is $\SU{2}$-equivariant. Then, composing $\lambda\colon\U{1} \to \SU{N}/\ZZ_N$ with the adjoint representation, we obtain a $\U{1}$-representation on $\Endo{\CC^N}$. With our choice of $\Lambda$, the $\U{1}$-action can be made explicit,
\begin{align}
 \ad{\e^{\Lambda t}} \left(\begin{array}{cc} M_{11} & M_{12} \\ M_{21} & M_{22} \end{array}\right) = \left(\begin{array}{cc} M_{11} & \e^{\im t}M_{12} \\ \e^{-\im t}M_{21} & M_{22} \end{array}\right),
\end{align}
where 
\begin{align}
 &M_{11}\in\Endo{\CC^m}, \quad M_{12}\in\Hom{\CC^n,\CC^m}, \\
 &M_{21}\in\Hom{\CC^m,\CC^n}, \quad M_{22}\in\Endo{\CC^n}.
\end{align}
Since in our conventions $\exp(E_3 t)$ takes precisely one full turn in $\U{1}$ as $t$ runs from $0$ to $4\pi$ (see section \ref{sec:ansatz}), the space of matrices $M_{12}$ is a $\U{1}$-invariant subspace of weight $2$, the matrices $M_{21}$ form a subspace of weight $-2$, and the matrices $M_{11}$ and $M_{22}$ together form a subspace of weight $0$. We therefore have the decomposition
\begin{align}
 \curlE = \bigoplus_{l=2,0,-2} E_{l} \otimes \Ob{l},
\end{align}
where $\Ob{l}$ denotes the line bundle of degree $l$ over $S^2$, and the $E_l$ are complex vector bundles over $\Sigma$ with
\begin{align}
 \rk\, E_2 = \rk\, E_{-2} = mn, \quad \rk\, E_0 = m^2 + n^2.
\end{align}
The residual gauge group $\curlH$ from \eqref{eq:2d_H} acts on $\Endo{\CC^N}$ as
\begin{align}
 \ad{h} \left(\begin{array}{cc} M_{11} & M_{12} \\ M_{21} & M_{22} \end{array}\right) =
 \left(\begin{array}{cc} h^m\, M_{11} (h^m)^{-1} & h^m\, M_{12} (h^n)^{-1} \\ 
			 h^n\, M_{21} (h^m)^{-1} & h^n\, M_{22} (h^n)^{-1}\end{array}\right), \label{eq:H_action}
\end{align}
where, in analogy with the notation in \eqref{eq:2d_trans},
\begin{align}
 h = \left(\begin{array}{cc} h^m & 0 \\ 0 & h^n \end{array}\right) \e^\frac{2\pi\im k}{N}.
\end{align}
From \eqref{eq:H_action} one can deduce that $E_{-2} \cong E_2^*$, the dual bundle of $E_2$.

\begin{figure}[h] 
 \begin{center}
  \psfrag{homom}{$\Ad{\Phi}$}
  \psfrag{bun_2}{$E_{2}$}
  \psfrag{bun_0}{$E_{0}$}
  \psfrag{bun_-2}{$E_{-2}$}
  \includegraphics[scale=0.65]{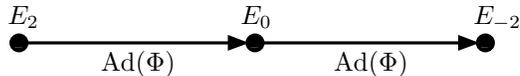}
  \caption{Quiver for the reduction with $\Lambda$.}
  \label{fig:quiver}
 \end{center}
\end{figure}

One also checks that $\Phi$ from \eqref{eq:little_higgs} acts on $\Endo{\CC^N}$ by the adjoint representation $\Ad{\Phi}=[\Phi,\:\:]$ as follows,
\begin{align}
 \Ad{\Phi} \left(\begin{array}{cc} M_{11} & M_{12} \\ M_{21} & M_{22} \end{array}\right) = 
 \left(\begin{array}{cc} -M_{12}\phi & 0 \\ \phi M_{11}-M_{22}\phi & \phi M_{12} \end{array}\right). \label{eq:Higgs_action}
\end{align}
Thus, $\Ad{\Phi}$ gives rise to homomorphisms between the bundles $E_2$, $E_0$, $E_{-2}$. This $\SU{2}$-equivariant set-up is captured by the quiver diagram in figure \ref{fig:quiver}.

We notice a new feature in the quiver in figure \ref{fig:quiver}, namely a redundancy in the quiver description: Both homomorphisms in figure \ref{fig:quiver} are determined by the same non-abelian Higgs field $\phi$, as can be seen from \eqref{eq:Higgs_action}, and $E_2$ and $E_{-2}$ are dual bundles. This redundancy stresses again that for Yang--Mills gauge group $\SU{N}/\ZZ_N$ it is more natural to work with principal bundles rather than vector bundles.

We conclude this section with a note on the generality of the dimensional reduction scheme we have employed in this paper. Since the results in \cite{Harnad} were obtained for rather general coset spaces, our analysis of equivariant principal bundles and connections should also generalize to coset spaces other than $S^2=\SU{2}/\U{1}$. In fact, a large amount of work has been carried out to study the quiver structures that arise from reductions over more general coset spaces, see for example \cite{Alvarez2, Lechtenfeld, Harland, Popov_double}. Most of the literature on this subject, however, focuses on vector bundles, and it is likely that for gauge groups other than $\U{N}$ or $\SU{N}$ a description in terms of principal bundles is more natural. As in this section, we expect that after reverting to the vector bundle language, the resulting quivers will be of rather special type and will carry redundant information.

\section{Conclusions} \label{sec:conclusions}

In the present paper we have further developed the idea that gauge theories describing vortices on a surface $\Sigma$ can be obtained from pure Yang--Mills theory on $\Sigma\times S^2$ by imposing spherical symmetry over $S^2$. Due to this symmetry, Higgs fields arise in the reduced theory on $\Sigma$, and the self-duality equations on $\Sigma\times S^2$ reduce to Bogomolny equations for vortices on $\Sigma$. Although various classes of such reductions have previously been studied in the literature, we have established a number of interesting properties of a special class of dimensional reductions: In this class the reduced theory contains a single, non-square, matrix-valued Higgs field, which is acted upon by gauge groups from the left and the right. Most notably, such a theory generally has a positive lower bound on its energy density, so the total energy is infinite unless $\Sigma$ has finite area. 

For a given area $A_{\Sigma}$ we have derived both upper and lower bounds on the energy and vortex number of solutions to the Bogomolny equations. Conversely, for a given vortex number, we have found the finite range of $A_{\Sigma}$ where solutions of the Bogomolny equations are possible. We have given examples of maximally abelian solutions, but these only exist for special values of $A_{\Sigma}$, and it remains an open problem to investigate genuinely non-abelian solutions. A special case of our construction gives the electroweak model with Bogomolny-type vortices investigated in \cite{BimonteLozano}, and we refer to one of our energy bounds as the \textit{Bimonte-Lozano} bound.

Furthermore, a bundle theoretic investigation has shown that while the reduced theory on $\Sigma$ has integral vortex number, the instanton number in the symmetric Yang--Mills theory on $\Sigma\times S^2$ is generally fractional. This is a consequence of choosing to work with the Yang--Mills gauge group $\SU{N}/\ZZ_N$, which is required by the geometry of $S^2$ if one wants to obtain a non-square, matrix-valued Higgs field. Further bundle theoretic considerations have led to an interpretation of our theories as rather special examples of quiver bundle constructions. The focus of future work could be on clarifying when the $\SU{N}/\ZZ_N$ gauge theory with the imposed symmetry can be lifted to a theory with gauge group $\SU{N}$ which still has the same symmetry. When this is the case, our analysis can be carried out entirely in terms of vector bundles, and relations to existing results can be expected. A somewhat opposite direction to explore is the study of Yang--Mills theories with other gauge groups that do not admit a natural description in terms of vector bundles.

\section*{Acknowledgements}

NSM thanks Giuseppe Marmo and the Theory Group, INFN, Naples for hospitality, and thanks Giuseppe Bimonte for drawing attention to ref.~\cite{BimonteLozano}. Both authors wish to thank Norisuke Sakai for useful discussions. NAR thanks Julian V. S. Holstein for several discussions on bundles. NAR is financially supported by EPSRC, the Cambridge European Trust, and St.~John's College, Cambridge.

\end{document}